\documentstyle[prd,aps,floats,epsf]{revtex}

\renewcommand{\thefootnote}{\fnsymbol{footnote}}
\def\be{\begin{equation}}
\def\ee{\end{equation}}

\def\ea{\end{eqnarray}}
\newcommand\pubnumber{SLAC-PUB-9653\\ BROWN-HET-1346}
\newcommand\pubdate{\today}
\newcommand\hepnumber{hep-th/0302160}

\def\SLAC{Stanford Linear Accelerator Center and ITP\\
    Stanford University, Stanford, California 94309 USA}
\def\doeack{\footnote{Work supported by the Department of Energy,
                     contract DE--AC03--76SF00515.}}
\def\Brown{ Department of Physics, Brown University, Providence, RI 02912, USA  }
\def\UBC{Department of Physics and Astronomy, University of British Columbia,\\
Vancouver, BC, V6T 1Z1, CANADA }

\def\Title#1{\begin{center} {\Large #1 } \end{center}}
\def\Author#1{\begin{center}{ \sc #1} \end{center}}
\def\Address#1{\begin{center}{ \it #1} \end{center}}
\def\andauth{\begin{center}{and} \end{center}}
\def\submit#1{\begin{center}Submitted to {\sl #1} \end{center}}
\newcommand\pubblock{\rightline{\begin{tabular}{l} \pubnumber\\
         \pubdate \\ \hepnumber \end{tabular}}}

\def\submit#1{\begin{center}Submitted to {\sl #1} \end{center}}

\begin{document}
\begin{titlepage}
\pubblock

\vfill
\Title{Non-Topological Inflation from Embedded Defects}
\vfill
\Author{Stephon Alexander\doeack}
\Address{\SLAC}
\medskip
\andauth
\medskip
\Author{Robert Brandenberger}
\Address{\Brown}
\medskip
\andauth
\medskip
\Author{Moshe Rozali}
\Address{\UBC}
\vfill

\vfill
\submit{Physical Review D}
\vfill
\end{titlepage}
\def\thefootnote{\fnsymbol{footnote}}
\setcounter{footnote}{0}
\tableofcontents
\newpage

\renewcommand{\topfraction}{0.99}
\renewcommand{\bottomfraction}{0.99}
\twocolumn[\hsize\textwidth\columnwidth\hsize\csname
@twocolumnfalse\endcsname

\title
{\Large {\bf  Non-Topological Inflation from Embedded Defects}}

\author{Stephon Alexander $^1$\footnote{e-mail: stephon@itp.stanford.edu},
Robert Brandenberger
$^2$\footnote{e-mail: rhb@het.brown.edu } and
Moshe Rozali $^3$\footnote{e-mail: rozali@physics.ubc.ca}
\\
$^1$ SLAC and Institute for Theoretical Physics, Stanford University,\\
Stanford, CA 94305, USA\\
$^2$ Department of Physics, Brown University, Providence, RI 02912, USA\\
$^3$ Department of Physics and Astronomy, University of British Columbia,\\
Vancouver, BC, V6T 1Z1, CANADA}

\date{\today}
\maketitle

\begin{abstract}

We discuss a new mechanism of obtaining a period of cosmological
inflation in the context of string theory. This mechanism is based
on embedded defects which form dynamically on higher dimensional
D-branes. Such defects generate topological inflation, but unlike
topological inflation from stable defects, here there is a natural
graceful exit from inflation: the decay of the embedded defect. We
demonstrate the idea in  the context of a brane-antibrane
annihilation process. The graceful exit mechanism suggested here
applies generically to all realizations of inflation on
D-branes.

\end{abstract}

]

\vskip 0.4cm

\section{Introduction}

There has been a lot of interest recently in exploring ways of
obtaining a period of cosmological inflation from string theory. A
realization of inflation in string theory should  address some of
the conceptual problems of quantum field theory realizations of
inflationary cosmology \cite{RHBrev}. Since string theory contains
many moduli fields which are massless at the perturbative level,
stringy inflation may yield a natural mechanism to produce
fluctuations of the required small amplitude \cite{lyth}. A
description of inflation in the context of string theory would
also automatically resolve the {\it trans-Planckian problem}
\cite{MB00} of inflationary cosmology, since it would yield a
complete description of the dynamics of fluctuations during the
entire inflationary period. Furthermore, since one of the goals of
string theory is to provide a nonsingular cosmology, a stringy
inflation model should also allow one to resolve the singularity
problem \cite{Borde} of inflationary cosmology.

Recent developments in particle phenomenology have explored the
possibility that the  our matter fields are confined to  a
four-dimensional space-time hypersurface in a higher-dimensional
bulk space-time \cite{add,rs}. In the context of string theory
this hypersurface arises naturally as a  D-brane, one of possibly
many branes that reside in the bulk spacetime \cite{ABE}. However,
for our purposes here,  a sufficient starting point  is some
higher dimensional field theory (coupled weakly to gravity). This
higher dimensional theory can be completed in the ultraviolet by
embedding in critical string theory, but also by using the $(2,0)$
fixed point \cite{seven}, little string theory
\cite{LST,Seiberg:1997zk}, or by deconstruction \cite{deco}. In
order to have a  definite mental picture we assume the higher
dimensional theory is embedded in string theory, and is realized
on D-branes.

This ``brane-world" scenario   has generated new realizations of
cosmic inflation. In one setup \cite{braneinfl,Burgess}, the
inflaton is the separation between a brane-antibrane pair.
However, it has been shown \cite{Burgess,BGW} that in this context
it is not easy to obtain initial conditions which lead to a
sufficient period of inflation. Another possibility is to have
inflation generated by branes at an angle \cite{angle}, or by a
configuration of defects of different dimensionalities
\cite{Kallosh}. This has the advantage of possibly providing a
mechanism for localization of chiral matter on the inflating
locus.

An alternative way to obtain inflation from brane collisions was
suggested in \cite{Stephon}, based on the realization
\cite{defect} that topological defects of co-dimension greater or
equal to one are formed on the brane worldvolume during the
collision of a brane-antibrane pair. In the simplest example, the
order parameter of the phase transition which occurs when the
brane-antibrane pair meets is given by a complex tachyon field
\cite{tachyon} which condenses at a nonvanishing expectation
value. The vacuum manifold of the tachyon field ground state
values has the topology of $S^1$. By causality \cite{Kibble}, the
values of the condensate in the vacuum manifold are uncorrelated
on large length scales, and hence a network of co-dimension 2
defects inevitably will form. In the case of a $D5$-${\bar D}5$
brane pair, the resulting topological defect is a $D3$ brane which
could be our world. Provided that the thickness of the topological
defect is larger than the Hubble radius, the defect core can
undergo inflation ({\it topological inflation} \cite{topolinfl}).

A problem with the mechanism of \cite{Stephon} (and of models of 4
 dimensional topological inflation \cite{topolinfl}) is the {\it
graceful exit problem}: how does inflation end, and a transition
to the usual late time cosmological evolution occur?

In this paper, we provide a simple solution to this problem, which
generalizes to all above mentioned ``brane-world'' models of
inflation. We point out that the defects which form in brane
collisions will often be {\it embedded defects} (which are non-BPS
branes in the string context) (see \cite{Nambu,embed} for a
discussion of embedded defects in field theory). These embedded
defects can be stabilized at high temperatures by plasma effects
\cite{Nagasawa1,Nagasawa2}, but decay at a sufficiently low
temperature, thus providing a natural graceful exit mechanism from
inflation. In the context of field theory inflation, such a
scenario was suggested in \cite{Lepora:1996ue}.

We may call the proposed inflationary model ``Higher Dimensional
(Non-) Topological Inflation'' since inflation is taking place in
the higher-dimensional   brane worldvolume, not only in the
directions of the defect (which are the dimensions of our
four-dimensional physical space-time).  Assuming that there is a
mechanism to stabilize the radion in the two (compact)
distinguished extra dimensions (the directions of the original
brane orthogoal to the defect), we would have found a way of
making {\bf two} of the internal dimensions larger than the
others, thus making contact to the scenario of \cite{ADD}
\footnote{Note that one also obtains two internal dimensions
larger than the others in the ``dynamical de-compactification''
scenario resulting from brane gas cosmology \cite{BV,ABE}.
However, in that context there is no reason to expect a large
hierarchy of scales.}. Alternatively, the two extra dimensions may
be taken to be infinite; as one moves away from the core of the
defect, inflation proceeds more slowly, and ends earlier. Thus the
resulting spacetime is expected to be warped, making contact with
the work of \cite{rs}. The question of which alternative is
realized in specific examples needs a detailed, model dependent
analysis, which we hope to return to in the future.

The causality argument (Kibble mechanism \cite{Kibble}) which
ensures that in a phase transition leading to the possibility of
topological defect formation such defects inevitably will form,
with a separation smaller or comparable to the Hubble length, also
ensures the formation of stabilized embedded defects in models
which admit them, as will be explained later. Thus, there is no
initial condition problem for our model of inflation.

In field theory models with a symmetry breaking scale $\eta$ much
smaller than the Planck scale, the width of a defect is typically
much smaller than the Hubble length. Thus, topological inflation
will not occur. An advantage of inflation in the context of higher
dimensions is that there is a new scale, the brane tension, which
is naturally higher than the Planck mass scale \footnote{This is a
standard useful property of D-branes: they can be very heavy
without causing large back-reaction.}.
 As will be shown in Section
III, in this context the defect width can easily be larger than
the Hubble length, thus making defect-driven inflation possible.

Our proposed mechanism of topological inflation from embedded
defects can apply in situations more general than that of a
brane-antibrane annihilation process. For example, in the context
of brane intersections (the intersection region undergoing
topological inflation), the branes may partially unwind around
each other, leaving behind some non-inflating components. This
process excites the off-diagonal open strings, which have no
geometrical interpretation.

 The
outline of this paper is as follows. In the next section, we
discuss a particular example in which one can realize topological
inflation from an embedded defect, namely embedded defects forming
in a brane-antibrane annihilation process. Next, we discuss how
topological inflation from embedded defects has a natural graceful
exit mechanism. Finally. we summarize our results, and discuss
further applications of the basic idea of topological inflation
from embedded defects in the context of brane physics.

\section{Embedded Defects from Higher Dimensional D-branes}

We will start with a brief review of the original scenario of
\cite{Stephon}. The scenario of \cite{Stephon} takes as a starting
point  a pair of parallel $D5$ and ${\bar D}5$ branes approaching
each other. As studied in \cite{tachyon}, when the pair gets
sufficiently close, a tachyon instability sets in. For minimal
gauge content of the branes, the tachyon can be described by a
complex scalar field $\phi$ with a standard symmetry breaking
({\it Mexican hat}) potential
\be \label{pot1} V(\phi) \, = \,
\lambda \bigl( |\phi|^2 - \eta^2 \bigr)^2 \, ,
\ee
where the point
$\phi =0$ corresponds to branes at zero separation, and
 $\eta$ is the symmetry
breaking scale. The vacuum at $|\phi| = \eta$ corresponds to the
closed string vacuum, where the original branes have annihilated.

The vacuum manifold of the tachyon condensates is $S^1$, and hence
according to the usual Kibble argument \cite{Kibble}, the fact
that the condensate values must be uncorrelated on scales larger
than the Hubble radius inevitably leads to the formation of stable
co-dimension 2 defects, in this case $D3$ branes, on the
worldvolume of the original $D5$ branes. Unlike most of the
discussion of defects on annihilating D-branes, we are interested
in the case where the resulting D3 brane is not BPS saturated, so
that inflation is possible.

In terms of gauge theory on the original brane worldvolume, the
above process corresponds to the symmetry breaking \be U(1) \times
U(1) \, \rightarrow \, U(1) \, , \ee where one of the $U(1)$
factors on the left hand side of the equation lives on either of
the branes, and the unbroken subgroup corresponds to the diagonal
$U(1)$ factor. The standard homotopy arguments (see e.g.
\cite{toprevs} for a discussion of such arguments in the context
of topological defects) yield
\be \Pi_1({\cal M}) \equiv
\Pi_1(G/H) \, = \, {\cal Z} \, ,
\ee
where ${\cal M}$ is the
vacuum manifold, and $G$ and $H$ are the full gauge group and the residual
gauge group after symmetry breaking, respectively
\footnote{In the usual discussion this
topological charge is the central charge appearing in the
supersymmetry algebra, and the resulting defect is then BPS
saturated. For inflation to occur, this charge should not be that
central charge, and the defect should be non-BPS. In any event,
the embedded defects discussed later are always non-BPS.}.

This set-up is not restricted to annihilating D-branes. Suppose we
start with any higher dimensional D-brane configuration, and that
at initial times the set-up is  removed from its vacuum. This
vacuum may be the closed string vacuum, but also an open string
vacuum, where the branes align in a supersymmetric fashion. In the
process of relaxation to the vacuum, the Kibble mechanism will
ensure the creation of local defects, for example brane
intersections. It is those defects which seed inflation in the
higher-dimensional topological context, and which are the subject
of our discussion. In particular, the energy scale associated with
this initial phase transition can be much smaller than the string
scale, so the above (higher dimensional) field theory
considerations apply.

 Our
idea, in this context, is to concentrate on the case of embedded,
non-topological, defects. For this end we  enhance the gauge
symmetry on each of the branes, e.g. from $U(1)$ to $SU(2)$.
Indeed, this arises naturally in the context of fivebranes in type
I theory\footnote{Alternatively,  our D-brane set-up can be placed
at a $Z_2$ orbifold singularity \cite{DM}}. In this case the
worldvolume gauge group is  $SU(2) \times SU(2) = SO(4)$. The
tachyonic scalar fields  transform as $(2,2)$ under $SU(2) \times
SU(2)$\footnote{There are also other matter fields, which are
assumed, like all  moduli, to have been stabilized by a separate
mechanism.}.

Now, the symmetry breaking pattern during tachyon condensation
becomes \be \label{newsym} SU(2) \times SU(2) \, \rightarrow \,
SU(2) \, . \ee In this case, no stable topological defects (in
particular no co-dimension 2 defects) form during this phase
transition since the vacuum manifold is ${\cal M} = S^3$ and hence
$\Pi_1({\cal M}) =  1$. Note that in the context of studies of
branes in Type I string theory it is already known that the $D3$
brane is unstable \cite{Uranga}.

The situation is similar to what happens in the electroweak theory
where (for vanishing Weinberg angle), the symmetry breaking is
$SU(2) \rightarrow 1$, as effectively in the above case
(\ref{newsym}). However, as is well known from analysis of the
electroweak theory \cite{embed}, it is possible to construct {\it
embedded defects}, solutions of the equations of motion which
correspond to unstable defects. In the case of the electroweak
theory in four space-time dimensions, the defects are {\it
electroweak strings}, an example of embedded strings
\footnote{Another example of embedded strings occurs in the sigma
model description of low energy QCD in the limit of vanishing bare
quark masses \cite{Carter}. In this case, the embedded strings are
the pion strings \cite{pion}.}.

As was recently realized \cite{Nagasawa1}, such embedded defects
can be stabilized at finite density by plasma effects. Consider
again first the example of the standard electroweak theory. The
order parameter of the symmetry breaking phase transition has four
real components, two of which are electrically charged, two are
electrically neutral. In the presence of a thermal bath of photons,
the effective potential will then be lifted in the charged scalar
field directions by more than it is lifted in the neutral scalar
field directions. Thus, the vacuum manifold corresponding to the
theory in a photon plasma is $S^1$ and not $S^3$. This can
stabilize a subset of embedded strings, namely the string
configurations constructed from the neutral scalar fields (these
are the electroweak Z-strings in the case of the standard
electroweak theory).

Note that if the embedded defects are thermally stabilized
at the temperature just below the phase transition temperature,
then the causality argument (Kibble mechanism \cite{Kibble})
applies and states that at least of the order one such defect
will form per correlation volume (which in turn is smaller than
the Hubble volume). The argument is as follows: during the
phase transition, the order parameter relaxes to the vacuum
manifold $S^3$. However, in the presence of thermal stabilizing effects
it will relax to the reduced vacuum manifold $S^1$. However, which
value in $S^1$ is taken on is random on scales larger than the
correlation length. Hence, defects will form. This process was
studied in detail in the case of semilocal strings (related to
embedded strings) in \cite{semilocal}.

Our idea is to apply this mechanism to brane cosmology. We need to
assume that the branes contain a thermal bath of gauge
fields which couple in the same way to the $SU(2)$ symmetry
breaking field of our
brane setup as the photon does to the $SU(2)$ symmetry
breaking order parameter in the electroweak theory. Ideally,
the usual photon field could play
this role. In this case, it would be natural to have only
this field in thermal equilibrium at late times, since it would
be the only massless gauge field. The
situation would then be identical to the electroweak theory discussed
above. Given the above assumptions about field content and coupling
to the $SU(2)$ symmetry breaking order parameter, it is completely
natural from the point of view of cosmology (hot initial state in
the very early Universe) to have the field excited in the required
way.

In this case, the phase transition corresponding to brane
collision will, by the Kibble mechanism \cite{Kibble}, inevitably
produce a network of co-dimension 2 branes. These defects will
remain stable until the matter density on the branes has
sufficiently diluted.

Let us now be a bit more specific. Given an $SU(2)$ gauge group
on each of the $D5$ branes, the world volume theory of a
coincident $D5$-${\bar D}5$ brane pair is $SU(2) \times SU(2)$,
and any gauge field configuration can be written as
a $4 \times 4$ matrix of complex entries, the upper diagonal
$2 \times 2$ entries corresponding to a $SU(2)$ matrix
describing open strings which start
and end on the first brane, the lower diagonal $2 \times 2$ entries
describing a $SU(2)$ matrix corresponding to the open strings
starting and ending on the other brane. The two off-diagonal
$2 \times 2$ sub-matrices are $SU(2)$ matrices corresponding to
open strings beginning and ending on different branes. It is
these latter sectors which contain the tachyon \cite{Sen:1999mg}.
In the case of $U(1)$ gauge fields on each of the branes, the
tachyon is a complex scalar field (one real component from
each of the sectors), in the case of $SU(2)$ gauge fields, there
are two real components in each sector.

Let us now assume that on each of the branes there is a thermal
bath of a $U(1)$ subgroup of $SU(2)$, and the other gauge
fields assumed to be vanishing. In this case, the gauge fields
excited on the branes will only couple to one of the two real
tachyon fields in each sector. Let us combine these two
fields into a complex field $\chi$ (charged with respect
to the excited gauge field), the other two into a
complex field $\phi$ (neutral with respect to the excited gauge
field). The effective world volume Lagrangian
then becomes
\be
{\cal L}_{\rm eff} \, = \, - {1 \over 4} F^{\mu \nu}F_{\mu \nu} +
{1 \over 2} {\bar D_{\mu} \chi} D^{\mu} \chi +
{1 \over 2} \partial_{\mu} {\bar \phi} \partial^{\mu} \phi - V(\phi, \chi) \,
\ee
where the $D_{\mu}$ stands for the gauge covariant derivative,
the gauge field being the one excited on the brane. Hence, the
gauge field directly couples only to the charged scalar field
$\chi$ and not to the neutral field $\phi$. The field strength
tensor of the gauge field $A_{\mu}$ excited on the brane is
denoted by $F_{\mu \nu}$, but the $F^2$ term in the Lagrangian
will not be important in the following. The tachyon potential $V$
has the typical symmetry breaking form
\be
V(\phi, \chi) \equiv
V_0(\phi, \chi) \, = \, \lambda \bigl( |\phi|^2 + |\chi|^2 - \eta^2 \bigr)^2
\, ,
\ee
where $\lambda$ is a coupling constant, and $\eta$ is the ground state
expectation value of the tachyon field magnitude. The corresponding
vacuum manifold is $S^3$.

The above effective Lagrangian is the same which describes
the sigma model of low energy QCD in the chiral limit, coupled
to an external bath of photons \cite{Nagasawa1,Carter}.
It also
describes the standard electroweak theory with only the photon
field excited \cite{Nagasawa2}. In the presence of a thermal bath
of $A_{\mu}$ gauge fields, we can take the thermal average of
the Lagrangian. Terms linear in $A_{\mu}$ vanish, the quadratic
term $A_{\mu} A^{\mu}$ becomes $\kappa T^2$, where $T$ is the
temperature of the bath. Thus, an effective mass term for the
$\chi$ field is generated. The effective potential becomes
\be
V(\phi, \chi) \, = \, V_0(\phi, \chi) + g^2 \kappa T^2 |\chi|^2 \, .
\ee
Hence, the vacuum manifold is lifted in the charged scalar field
directions. The vacuum manifold of the effective potential becomes $S^1$.
Thus, co-dimension 2 defects in which the neutral scalar field
configuration takes on the Nielsen-Olesen \cite{Nielsen:cs}
form become important as  metastable defects.

\section{Embedded Defects, Inflation, and Graceful Exit}

The idea of topological inflation is quite simple
\cite{topolinfl}: provided that the core of the defect is
comparable or larger than the Hubble radius, then the potential
energy density in the defect core will be larger than the tension
energy, and thus via standard minimal coupling to gravity, the
core will commence inflationary expansion. For defects associates
with an energy scale much smaller than the Planck scale, the width
is much too small to obtain topological inflation. However, if the
tension of the defect is given by an energy scale $\eta$ which
is comparable of larger than $m_{\rm pl}$, then the condition for
topological inflation is satisfied. A quantitative analysis
\cite{Sakai:1995nh} yields as condition for topological inflation
\be \eta \, > \, {1 \over 4} m_{\rm pl} \, . \ee

Let us give a brief derivation of this result. For a typical
symmetry breaking potential of the form (\ref{pot1}), the defect
width $w$ is of the order \cite{Nielsen:cs} \be w \, \simeq \,
\lambda^{-1/2} \eta^{-1} \, , \ee which can be seen by balancing
potential and tension energies. In the core, the potential energy
density dominates, in the outside regions of the defect the
tension energy density is larger. The condition for defect
inflation is that the core width is greater than the Hubble length
$H^{-1}$. Applying the Friedmann equation \be H^2 \, = \, {{8 \pi}
\over 3} G \rho \ee (where $\rho$ is the energy density) to the
defect core (where $\rho \simeq  V(\phi = 0)$), the condition
becomes \be \label{cond} {{\eta} \over {M_{\rm pl}}} \, > {1 \over
{2 \sqrt{2}}} \, . \ee Thus, for values of $\eta$ comparable or
larger than $m_{\rm pl}$ but coupling constant $\lambda \ll 1$,
both the condition (\ref{cond}) for topological inflation and the
condition $V(0) < m_{\rm pl}^4$ for applicability of the Friedmann
equations for the evolution of the background space-time are
satisfied.

The above conditions can easily be satisfied for defect arising on
D-branes in superstring theory. In this context, it has been shown
that the tachyon potential, computed in superstring field theory,
has the form \cite{Berkovits} (see also \cite{Kraus}) \be
\label{pot3} V(T, {\bar T}) \, = \, \tau (\alpha^{'})^2 \bigl(
|T|^2 - (\alpha^{'})^{-1} \bigr)^2 \, , \ee where $\tau$ is
proportional to the brane tension and $\alpha^{'}$ is the string
tension. Thus, if $\alpha' < m_{\rm pl}^2$, then the conditions
for topological inflation are satisfied. This condition requires
the (six dimensional) string coupling to be small.

Topological inflation from stable defects suffers from a graceful
exit problem. Inflation in the defect core continues forever, and
in order to make a successful transition to late time cosmology
one usually has to postulate that our region of space originated
from a region close to the original defect boundary. This is
already problematic in the case of field theory inflation in
 four dimensional space-time. Here the problem becomes  even more
acute, since  in
the context of brane world models, our matter fields have to be
localized on the locus that is to become our physical spacetime.

However, if the inflationary expansion takes place in the core of
an embedded defect, the graceful exit problem is naturally resolved:
after some finite time, the defect decays, the potential energy
density disappears, and inflation stops.

As was shown in \cite{Carter}, embedded defects typically acquire
superconducting current. These currents help stabilize the
embedded defects down to very low temperatures. At the temperature
$T_c$ below which the thermal barrier becomes too low to stabilize
the embedded defect, a core phase transition \cite{core} (see also
\cite{Brito}) occurs
(the charged scalar fields acquire nonvanishing expectation
values), but the defect persists \cite{Nagasawa2}. The defect
decays only when the gauge field responsible for the stabilization
of the defect ceases to be in thermal equilibrium. Let us assume
that it is the photon which is responsible for the stabilization
of our embedded $D3$ brane. In this case, the decay temperature
$T_d$ could be \footnote{However, this is
an optimistic assumption since we are considering an epoch during
which the Universe is exponentially expanding - we thank
J. Khoury, W. Kinney and T. Baltz for a discussion on this point.}
as low as
\be T_d \, \sim \, 10^{-10} {\rm GeV}
\, ,
\ee
the temperature of recombination \footnote{Note that after decay,
the Universe will reheat to a very high temperature as long as
some of the energy of the embedded defect goes into standard
particle physics model matter.}.
If the temperature at
the onset of inflation is about $10^{17} {\rm GeV}$, in light of
the formula (\ref{pot3}) for the tachyon potential in  string
field theory quite a reasonable value (it is also the upper bound
on the scale of inflation in order not to overproduce
gravitational waves), then we (optimistically) obtain about 62 e-foldings of
inflation, only slightly larger than the minimum number of
e-foldings required \cite{Guth} for inflation to solve the horizon
and flatness problems of standard cosmology. Note that for this
small number of e-foldings, the physical wavelength of comoving
scales which correspond to present day cosmological scales was
larger than the Planck length at the beginning of inflation. Thus,
there is no trans-Planckian problem \cite{MB00} in this model.

As mentioned in the Introduction, in this mechanism of topological
inflation from embedded defects produced during $5$ and ${\bar 5}$
brane annihilations, two of the internal dimensions (namely those
parallel to the worldvolume of the initial branes) also inflate.
Given the low number of e-foldings of inflation which result in
this scenario, the size of the two large extra dimensions is
comfortably below the observational bounds (assuming that there is
a radion stabilization mechanism which sets after inflation in for
these directions).

\section{Summary and Discussion}

In this paper we have proposed a new way of obtaining inflation in
the context of brane physics. The mechanism is based on inflation
taking place in the core of an embedded defect. The embedded
defect is stabilized at high temperatures by plasma effects, as
studied in \cite{Nagasawa1}, but becomes unstable and decays at
lower temperatures, thus providing a natural graceful exit
mechanism from inflation.

We have suggested a specific realization of this idea in the
context of a brane-antibrane setup, where our world volume
corresponds to that of an embedded 3 brane which is generated in
the process of tachyon condensation of an annihilating 5 brane -
antibrane pair. To obtain an embedded defect in this setup, the
gauge groups on the 5 branes must be enhanced from the minimal
gauge group $U(1)$.

Clearly this idea is  more general than the brane-antibrane setup.
For example, in the context of brane intersections
\cite{Berkooz:1996km} an embedded defect can  be realized if the
gauge content on the branes is enlarged.   Some intersections can
unwind as the temperature cools, a process described in the field
theory as the decay of the embedded defect, and requires one to
excite  off-diagonal, non-geometrical open strings (see for
example \cite{joe} for a detailed discussion).

 More generally, it is well-known, at least in the BPS sector, that
stability of defects depends on parameters of the theory (see e.g.
\cite{Seiberg:1994rs}). This is well controlled for supersymmetric
states, by using the BPS formula, but is expected to be the case
generally. In this case, one can use the topological inflation
idea, and provide a natural graceful exit mechanism along the
lines of this paper. In this scenario the initial stable defect
causes inflation. As inflation proceeds, and effective parameters
of the theory change in response to cooling down, causing the
initial defect to become unstable. This paper is a concrete
example of this more general scenario.

Obviously, our scenario of ``topological inflation'' from an
embedded defect can also be realized in ordinary four dimensional
quantum field theory. However, in that context it appears
unnatural to obtain defects with the width required for
topological inflation, unless field values larger than the Planck
scale are invoked. As we have seen, in the context of the brane
setup, the existence of new physical scales (e.g. the brane
tensions) makes it easy to obtain defects of sufficient width to
support topological inflation.

We have also seen that in our realization of embedded defect
inflation in the context of an annihilating $5$ and ${\bar 5}$
brane pair, inflation can produce a hierarchy in the sizes of the
internal dimensions (or a warped geometry), since the two extra
spatial dimensions parallel to the worldvolume of the $5$ branes
also expand exponentially.

Recently, it was argued that de Sitter space vacua in string
theory will suffer from conceptual problems \cite{Dyson1,Dyson2},
if the lifetime of de Sitter space lasts longer than the
Poincar\'e recurrence time $t_{r} \sim e^{S_{0}}$, where $S_{0}$
is the entropy of the space. For example,  Kachru et. al provided
a stringy realization of a meta-stable de Sitter space which
tunnels to flat space-time in a timescale shorter than
$t_{r}$\cite{Kachru}. Our scenario is more reminiscent of the new
inflationary models, in that no tunneling is required to stop
inflation. Consequently our inflationary spacetime turns  into a
hot FRW space-time after $\sim 62$ e-foldings, a time scale
exponentially shorter than $t_r$.

\vspace{1cm}

{\bf Acknowledgments}

We are grateful to M. Berkooz, C. Burgess, K. Dasgupta, A.-C.
Davis and G. Gabadadze for stimulating discussions. One of us (RB)
wishes to thank E. Baltz, J. Khoury and W. Kinney for useful
comments. This work was supported in part by the U.S. Department
of Energy under Contracts DE-FG02-91ER40688, TASK A (Brown
University) and DE-AC03-76SF00515 (SLAC), and by the Canadian
NSERC and the PIMS string theory CRG (at UBC). M.R thanks the
theory group  at Stanford  and the organizers of the APCTP-KIAS
winter school for hospitality while this work was completed.


\end{document}